# ECONOPHYSICS APPROACH AND MODEL ON MIXED ECONOMY

Ion SPÂNULESCU[*], Anca GHEORGHIU[*]

***Abstract.*** *In this paper the general principles and categories of mixed economy that currently exist in almost all countries of the world are presented. The paper also presents an Advanced Model of Mixed Economy with Threshold (AMMET), which is characterized by a reduced value (approx. 10-15%) of the State and public sector participation in the national economy and proposes and analyzes an econophysics model for the mixed economy.*

## 1. Introduction

The mixed economy system is, in essence, a system of market economy where, in addition to private agents and firms, the **State** also participates playing a role of coordination and control or direct participation in the economic process through the its public sector.

In the nowadays, in the case of a sharp increase of the State's role and bureaucracy, including of the public sector in the economy, the Keynes's equation $Y = C + I$ of the economic equilibrium for a closed economy and without the participation of the State [2] must be completed by the term $G$ which signifies the governmental and public expenditures together, namely:

$$Y = C + I + G, \qquad (1)$$

Equation (1) can be considered as the economic equilibrium equation for the case of Mixed Market Economy [1,4]. In this equation, the term $Y$ represents the total income at macroeconomic level, $C$ is global consumption, $I$ representing total investments, and $G$ – public and governmental spending [1,4].

[*] Hyperion University, Calea Călăraşilor 169, Bucharest, 030615

In this paper an econophysics model for mixed economy is presented and analyzed.

## 2. The Advanced Model
## of the Mixed Economy with Threshold (AMMET)

Currently the mixed economy has developed in almost all countries of the world, where there is a market economy (with the exception of a few socialist countries or totalitarian regimes). As shall be seen below, this model knows a great diversity depending on the degree of involvement of the State (or government) and the public sector in the economy.

When most of the states were formed, and their economies developed strongly, especially after the industrialization era, there were several economic doctrines, currents, or schools that promoted or supported different economic theories, the majority being contradictory and tainted by political opinions or specific economic interests etc. In terms of the mixed economy, the State's involvement in the economy here we can mention the existence of two currents that have developed since the second half of the 19$^{th}$ century, namely the Austrian School and the German School respectively [1,3,4].

The Austrian School emerged after 1870 and represented by Carl Menger, Friedrich von Wieser, Ingrid von Mises and others supported the noninterference of the state in economy, while the German School or the Rhine (or Renan) Model represented by Gustav von Schmoller, Werner Sombart and their disciples expresses the exact opposite views by giving an important role to the State's intervention in the market economy.

According to the theses promoted by the two mentioned schools, Michel Albert distinguishes two major categories of models of the mixed market economies currently in place, namely, the **Neo-American** or Anglo-Saxon model met in the U.S.A., England or Canada, and the **Rhine** (Renan) model represented by the economies of Germany, Switzerland, Netherlands and the Nordic countries (Sweden, Norway) [5]. The first model (Neo-American) is characterized by a minimum level of State intervention in the economy and a certain level of insurance of social protection measures, while the second model is characterized by a higher level (but not exaggerated, under 35-40%) of the public sector and a more significant intervention of the State in the socio-economic life of the country, as well as through a much higher degree of social protection (state pensions, unemployment benefits, free school education and health assistance etc.) [5-7].

To point out, that even in the case of the Renan Model the market has a dominant role, production and services are made especially in the private sector, and prices are adjusted on the basis of supply and demand under the laws of the free market.

Depending on the philosophy underlying the socio-economic structures in the mixed market economy countries and the degree of State involvement in the economy, the other mixed economy states belong to one or another of the two categories mentioned above or may present specific features such as Japan, China or France [1,3-7].

From the equation (1) it follows that the high value of public and government expenditure, noted together by the term *G*, strongly affects the efficiency of the investment process at a given *Y* income, resulting in either diminishing consumption *C* or investment *I* or their simultaneous decrease in different proportions, depending on the given economic conditions. Thus, from the analysis of the balance equation (1) for the mixed economy, it is obvious that in order to ensure optimum values for investment and consumption, the amount of non-productive expenditure *G* should be as low as possible, i.e. to establish a **minimum threshold** of interference of the State in the economy, which must be as small as possible, and a low value of the public sector in the national economy.

As a result, there is a high value diversity of the share of state and public sector participation in the Mixed Economy in different countries, ranging from values below 20-30% in more developed countries (USA, England etc.) and approaching the threshold of 45-50% in countries where a greater role is given to social protection policies or in which there is a control of the economy from state bodies (such as in France, Japan, China for example). Hence it follows that for full substantiation of the *Mixed Economy Model* is necessary to determine **the percentage** or the value of an **optimal threshold** of intervention and participation of the State and the public sector in the national economy or which are the percentage limits (lower and upper) that may vary the respective threshold values.

It is apparent from the foregoing that if we propose to adopt or improve the system of the mixed economy, then the **threshold** or limited **level** to which the State and public sector can intervene in the country's economy should be specified, otherwise the system can divert to the system of a centralized-type economy, which is not desirable. This complement that we propose to improve the system of the mixed economy gives the model the characteristics of an *"Advanced Model of the Mixed Economy with Threshold – AMMET"* – the name that we will continue to use when we refer to the Mixed Economy with a certain limited level of intervention of the public sector and the State in the economy [1,3,4].

However, the following question is asked: How big will the value of this threshold should be, in relation to the total income or gross domestic product? Considering the analysis and considerations made above, the value of this threshold must be as **small** as possible, i.e. the free private market economy sector must hold a comfortable majority, much higher than the State-controlled part.

In order to determine the optimum value of this threshold, in our studies we have resorted to an econophysics model for the mixed economy that will be presented in the next section, the preliminary results of this study being mentioned in our previous papers [1,4].

## 3. An Econophysics Approach and Model of Mixed Economy

For the full justification of the Econophysics Model of the Mixed Economy, the preliminary results obtained for the model were communicated to the International Conference on Econophysics, New Economy and Complexity – ENEC 2018, which took place at Hyperion University in Bucharest, Romania in May 2018 [1].

The econophysics model of mixed economy is based on the model of economic amplifier (Figure 1.a, b), which works by analogy with an electronic amplifier with solid state electronic devices (transistors or integrated analog circuits) (Figure 2.a,b) analyzed in our previous papers [8-11].

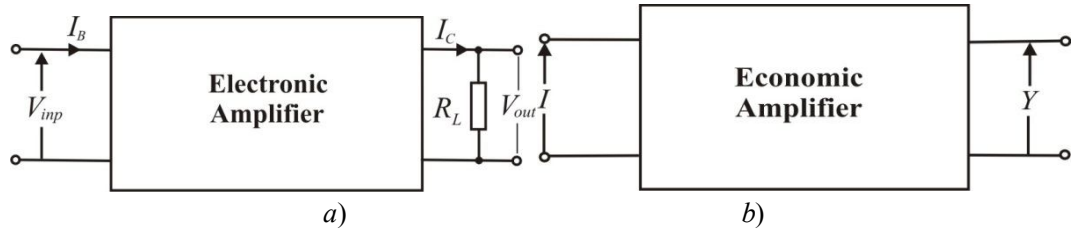

**Figure 1.** a) Simplified representation of an electronic amplifier with transistors or integrated circuits; b) The schematic representation of on economic amplifier.

The amplification factor of the single stage transistor amplifier in Figure 1.a in the presence of only polarization bias voltages applied to the transistor is basically given by the continuous current amplifier factor β of the bipolar transistor in common emitter configuration (EC), where emitter's electrode is **common** both for the input and output circuit (Fig. 2. a,b):

$$\beta = \frac{I_{out}}{I_{in}} = \frac{I_C}{I_B}. \tag{2}$$

In equation (2), $I_C$ represents the collector current at the output of the transistor from the amplifier with transistor or integrated circuits (Fig. 2a), and $I_B$ is the base current that is present in the input circuit [9-11]. In the papers [9, 10] is shown that similarly, an economic amplifier can be characterized by the amplification factor β $_{economic}$ given by [9-12]:

$$\beta_{ec} = \frac{Y}{I} \quad (3)$$

where *Y* is the aggregate income obtained on the basis of investments *I* from "the entry" of economic amplifier (Figure 1.b) [9-10].

We can imagine an econophysics model for mixed economy by analogy with the physical structure of a bipolar transistor (Fig. 2.b), which is the active amplification device from an electronic amplifier as the one in Figure 2.a, which – as we've shown in our previous paper [9-11] – perfectly shapes the economic development phenomenon in a sustainable economy (in which there is growth based on investments *I*) according to the equation (3) for the economic amplifier [9,10].

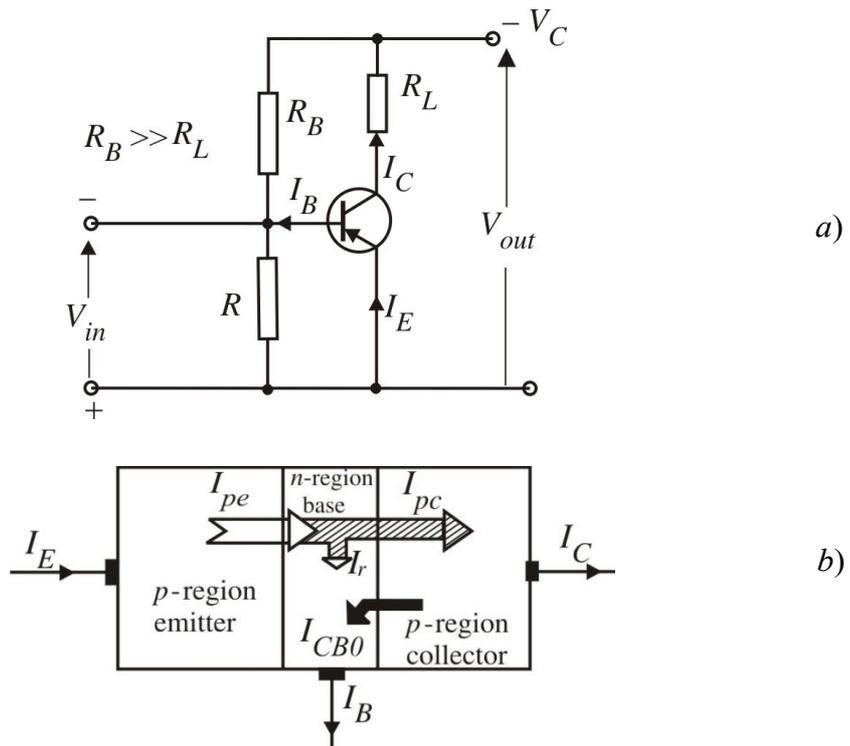

**Figure 2.** a) Schematic of electronic amplifier with a *p-n-p* transistors;
b) Schematic physical structure of a *p-n-p* transistor.

In the structure shown in Figure 2.b, the *p*-regions that represent the emitter and, respectively, the collector of the bipolar transistor with *p-n* junctions, are assimilated to the private sector of the country's market economy, while the *n* type intermediate region, referred to as the base of the transistor, represents the public sector, its participation and the degree of state intervention in the economy, being characterized by the value $I_B$ of the base current that closes through input circuit of the electronic amplifier (Figure 2.a) that shapes economic development through the associated economic amplifier. Thus, *G* factor of the economic equilibrium equation (1), showing the share of public and government expenditures is modeled by the $I_B$ current in the case of the electronic amplifier (Fig. 2.a,b) that shapes the economic amplifier, i.e. in the case of the econophysics model of the mixed economy instead of $I_B$ will be taken the *G* factor, so we can write:

$$I_B = G. \tag{4}$$

In equation (2) it is obvious that excessive increase of current $I_B$ is unacceptable, because it leads to drastic decrease or even cancellation of the amplification factor β, i.e. the economic amplifier doesn't work and economy stagnating. For the proper operation of the electronic amplifier and, respectively, of the mixed economy model, it is essential that $I_B$ and by default *G* to be as low as possible to get a reasonable value for the amplification factor β and the default $β_{ec}$.

From Electronic Physics it is known that the current $I_B$ is given, mainly, by the sum of the reverse current of electrons coming from the collector to the base, $I_{CBO}$ and the current $I_r$ of recombination in the base of positive charge carriers injected from emitter as a result of application of entrance signals and normal bias voltages of the amplification device (transistor) [4,13]:

$$I_B = I_r + I_{CBO}. \tag{5}$$

In equation (5) we see that for the reduction of the base circuit current $I_B$ is necessary to significantly decrease the width of the base that has the effect of significant decrease the level of recombination current $I_r$, the current $I_{CBO}$ having a constant and reduced value.

Decreasing the width of the base is equivalent to a reduction of the public sector and the state influence in economy, according to the econophysics model proposed for mixed economy. Correspondingly, as evidenced by the equation (5), decreasing the current $I_B$, respectively of the width of the base is equivalent to the significant reduction of public sector and state intervention in the economy, as well as the governmental expenditure from Equation (1). The proposed model has the great advantage that by using it we can also determine the **optimum threshold** or the maximum limit of the public sector and state intervention in the economy in order to have a performing and sustainable mixed economy.

Indeed, from those shown earlier, it appears that $I_B$ has a much lower value than the main current through the transistor, which is also the output current $I_C$ which allows for an amplified voltage $V_{out}$ at the terminals of a load resistance $R_L$ from the electronic amplifier (Fig. 2.a). As the value of the current $I_B$ is much smaller than the output current $I_C$ of the transistor in the electronic amplifier we can write that it represents a fraction $\alpha_{el}$ of the value of output current $I_C$ electronic amplifier, i.e. [4]:

$$I_B = \alpha_{el} I_C \qquad (6)$$

with $\alpha_{el}$ ranging between 0.1 and 0.9. Replacing (6) in equation (2) and assimilating the output current $I_C$ with the income $Y$ of the economic amplifier results for $\beta_{ec}$ of the electronic amplifier [4]:

$$\beta_{el} = \frac{I_C}{I_B} = \frac{I_C}{\alpha_{el} I_C} = \frac{1}{\alpha_{el}} \qquad (7)$$

and respectively:

$$I_B = \alpha_{el} I_C = \alpha_{el} Y. \qquad (6')$$

From equations (3) and (4) we can also write an equation similar to the equation (7), i.e.:

$$\beta_{ec} = \frac{Y}{G} = \frac{Y}{\alpha_{ec} Y} = \frac{1}{\alpha_{ec}}. \qquad (7')$$

Thus, the transistor's amplification can be increased by decreasing the base width as much as possible (consequently of the public sector in the economy in the case of econophysics model) which reduces the base current $I_B$ mainly by drastically reducing recombination current $I_r$ in the base region (see equation (5)). As a general rule, to compensate for some losses in the elements of an electronic amplifier circuit due to the influence of the environment (temperature, radiations) or bias voltages, in order to achieve the minimum amplification with the single stage amplifier (with one transistor) it is assumed that the current amplification factor β to be at least an order of magnitude larger than the unit, i.e. $\beta \simeq 10$. Using this condition and replacing the value of $\beta_{el} \simeq 10$ in equation (6), results for the fraction $\alpha_{el}$ a value of 0.1.

Although in practice $\beta_{ec}$ is much lower than $\beta_{el}$, for amplification, in the case of the economic amplifier can also choose a value $\beta_{ec}$ approximately equal to the value 10, as in the example above, for the electronic amplifier.

Assimilating the aggregate economic growth $Y$ with the country's Gross Domestic Product (GDP) and replacing $\alpha_{el} = \alpha_{ec} = 0.1Y$ in equation (6') results that the **optimal** percentage value for the degree of intervention in the economy of the state and the public sector given by the current $I_B$, respectively for the term $G$ of the econophysics model [4]:

$$I_B = 0.1\ Y = 0.1 GDP \qquad (8)$$

and, respectively:

$$G = I_B = 0.1\ GDP \qquad (9)$$

which represents the optimal value accepted for $G$, for optimal operation of mixed economy within the *Advanced Model of the Mixed Economy with Threshold – AMMET* presented in section 2 of the paper.

Taking into account the Equation (9) and the examples of the developed countries mentioned in section 2, for the significant reduction of public and governmental expenditure, $G$, it is proposed that the ideal (optimal) percentage of participation of the public sector and the interference of State in economics to be about 10% of the country's GDP [4].

In favor of adopting a small value (about 10% of GDP) for the public sector and the State intervention threshold in the economy, it also advocates the example of the mixed economy from highly developed countries and, in particular, the United States of America that have a small threshold of state interference in the economy [1,3,4].

The threshold of about 10% of GDP resulting for $G$ given by the equation (8) for the case when $\beta_{ec}$ equals 10 can be considered as an **optimum** threshold, and may be recommended to be used resulting a sustainable economic growth in mixed economy countries.

From our previous research [9-11] on the application of the electronic amplifier model in various situations of economic growth emerged that $β_{ec}$ for different applications acquires values between 1.95 and 6, so the value $β_{ec} = 10$ appears to be perfectly reasonable. In various practical situations, depending on economic and socio-political conditions specific to any country it is possible that the $β_{ec}$ to fluctuate between certain limits, taking values lower or higher than 10 taken in the example analyzed above. On the other hand, in all the considerations made on the various econophysics models, account must be taken of the complexity of the economic processes and laws in which, as a rule, the influence of the human factor is present, which can sometimes lead to unexpected results and therefore more difficult to control. Also, the laws of Physics – which is an exact science – cannot apply by going up to the identity between the Physical phenomena and the Economic laws analyzed, but only by **analogy** or **similarity** depending on the characteristics of the phenomenon or the investigated economic process. Thus, $β_{elec}$ for an electronic amplifier can take very high values, being of up to several orders of magnitude (approx. $10^3 – 10^6$ etc.) depending on the type and quality of the active electronic amplification devices used in the amplifier, while $β_{ec}$ can take values up to an order of magnitude (approx. 10) or lower (< 10) but which are perfectly normal for the economic amplifier adopted as an econophysics model [ 9-12].

In practice, depending on the political conceptions and social or economic conditions, the values for $α_{ec}$ or $β_{ec}$ may vary in much wider limits as seen from Figure 3. The diagram in Figure 3 may distinguish three distinct regions marked by areas A, B and C respectively.

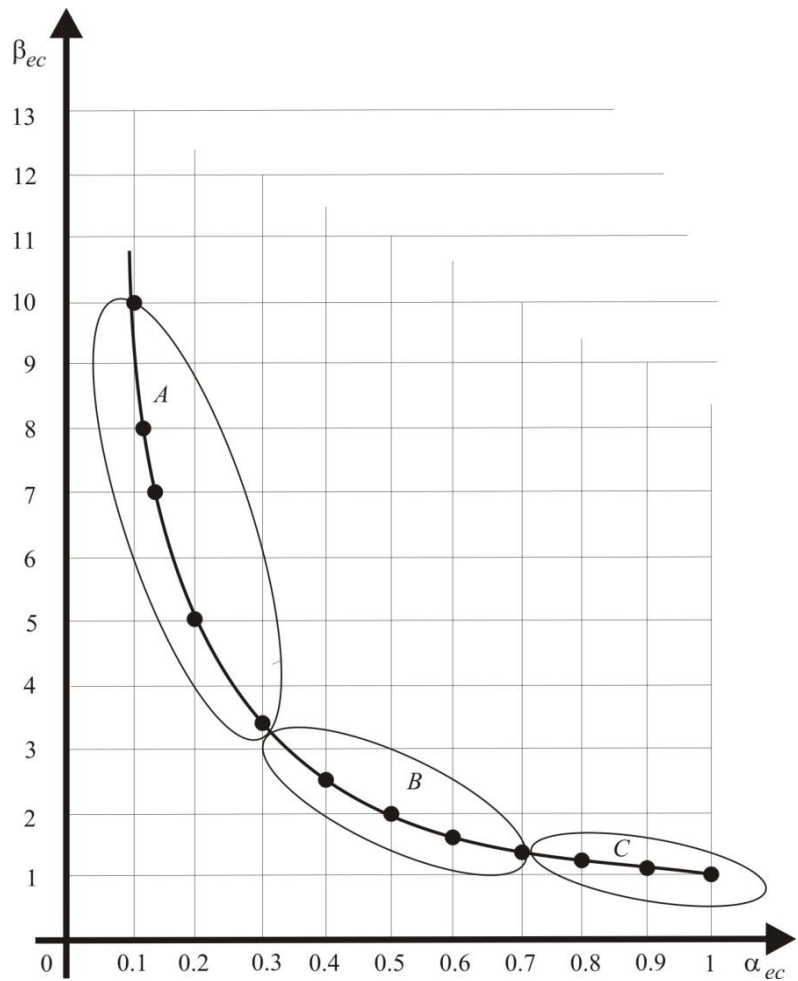

**Figure 3.** Graph of the $\beta_{ec}$ parameter values for different values of $\alpha_{ec}$.

Countries that adopt a sustainable mixed economy correspond to region A for values of the $\alpha_{ec}$ percentage between 0.085 and 0.35, while countries which, in addition to a harmonious economic development, give greater attention to effective social protection policies, correspond to region B for values of the $\alpha_{ec}$ percentage between 0.4 and 0.75; Region C for values of $\alpha_{ec}$ more than 0.75 corresponds to a weak economic growth (with $\beta_{ec}<1.5$), so it is not advisable to apply. It should be noted that the values proposed here for $\alpha_{ec}$ and $\beta_{ec}$ appear as a conclusion of the graphical representation based on the hypothesis of the electronic amplifier, and are not of an immutable nature. They may be adopted or not, according to economic and social policies and economic situations and economic reserves available to countries applying the mixed economy model analyzed in this work.

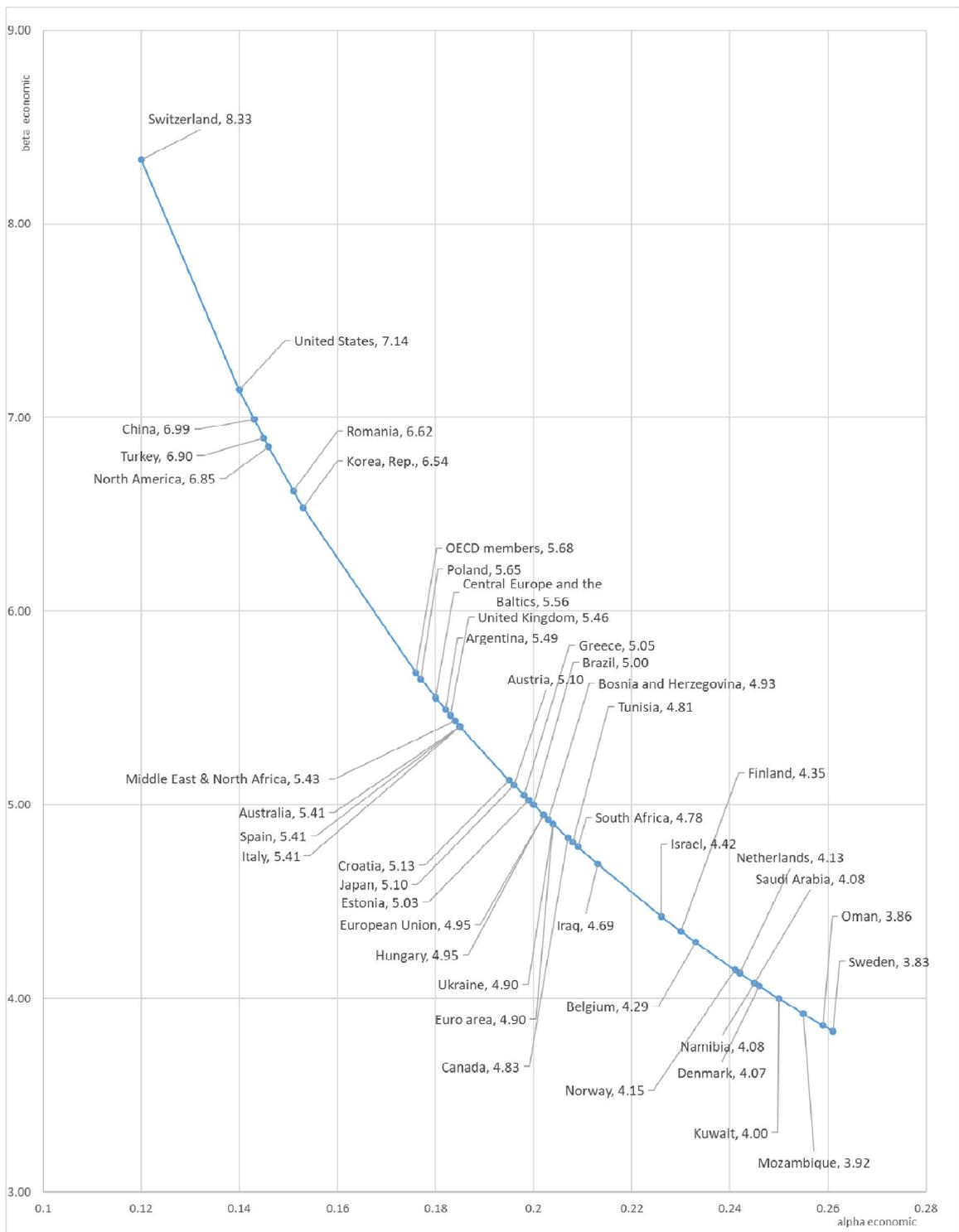

**Figure 4.** General government final consumption expenditure (% of GDP) represented by $\alpha_{ec}$ indicator and $\beta_{ec}$ values) for 2017 year. **Source:** https://data.worldbank.org

**Table 1**

| Country | α$_{ec}$ | β$_{ec}$ |
|---|---|---|
| Argentina | 0,182 | 5,494505 |
| Australia | 0,185 | 5,405405 |
| Austria | 0,196 | 5,102041 |
| Belgium | 0,233 | 4,291845 |
| Bosnia and Herzegovina | 0,203 | 4,926108 |
| Brazil | 0,2 | 5 |
| Canada | 0,207 | 4,830918 |
| Croatia | 0,195 | 5,128205 |
| Denmark | 0,246 | 4,065041 |
| Estonia | 0,199 | 5,025126 |
| Finland | 0,23 | 4,347826 |
| Greece | 0,198 | 5,050505 |
| Hungary | 0,202 | 4,950495 |
| Iraq | 0,213 | 4,694836 |
| Israel | 0,226 | 4,424779 |
| Italy | 0,185 | 5,405405 |
| Japan | 0,196 | 5,102041 |
| Korea, Rep. | 0,153 | 6,535948 |
| Kuwait | 0,25 | 4 |
| Mozambique | 0,255 | 3,921569 |
| Namibia | 0,245 | 4,081633 |
| Netherlands | 0,242 | 4,132231 |
| Norway | 0,241 | 4,149378 |
| Oman | 0,259 | 3,861004 |
| Poland | 0,177 | 5,649718 |
| Romania | 0,151 | 6,622517 |
| Saudi Arabia | 0,245 | 4,081633 |
| South Africa | 0,209 | 4,784689 |
| Spain | 0,185 | 5,405405 |
| Sweden | 0,261 | 3,831418 |
| Switzerland | 0,12 | 8,333333 |
| Turkey | 0,145 | 6,896552 |
| Tunisia | 0,208 | 4,807692 |
| Ukraine | 0,204 | 4,901961 |
| United Kingdom | 0,183 | 5,464481 |

| United States | 0,14 | 7,142857 |
| --- | --- | --- |
| Central Europe and the Baltics | 0,18 | 5,555556 |
| Euro area | 0,204 | 4,901961 |
| European Union | 0,202 | 4,950495 |
| Middle East & North Africa | 0,184 | 5,434783 |
| North America | 0,146 | 6,849315 |
| OECD members | 0,176 | 5,681818 |

**Source:** https://data.worldbank.org/indicator/ne.con.govt.zs

In Figure 4 and Table 1 above, $β_{ec}$ values are represented based on the $α_{ec}$, we mean according to the percentages listed in GDP for government spending without any mention the share of the public sector in economy. The values for $α_{ec}$ underlying the composition of the graph in Figure 4 are taken from the data of the World Bank at the level of the year 2017. The chart of Figure 4 confirm the results represented in Figure 3, composed on the basis of the econophysics model of the mixed economy presented in this paper, for the data between the years 2005-2007.

Compared with the 2005-2007 period, an increase in the number of countries that have adopted reduced values for $α_{ec}$ was observed, so implicitly values much lower for government spending $G = α_{ec}$ GDP, so these countries will enter area *A* of Figure 3, while the group of countries with high values for $α_{ec}$, respectively over 4.5, is drastically reduced, basically the *C* area from Figure 3 being inexistent in the figure 4.

The need to adopt a small percentage of the public sector and the smallest interference of the State in the economy does not appear only as a result of the application of the econophysics model presented here and the theoretical analysis of the equation (1) which – for reducing public expenditure (the term *G* of the equation (1) of the economic equilibrium), recommended a minimum participation of the state in the mixed economy – but also from the practical experience of the application of the minimum threshold model as is the case for countries with a strong economic development, which also confirms the validity of the proposed econophysics model for the mixed economy.

## 4. Conclusions

The system of mixed economy represents the market economy system where, in addition to economic agents and firms etc., the state also participates with coordination and control role, or can take part directly to the economic process through its public sector.

As has been shown in our previous works, the degree of state and public sector participation in the mixed economy can vary in quite a wide range by going from about 10% to nearly 90% depending on the political and social orientation in each country and the concrete economic conditions and trends of economic and social approach specific to those countries. In this situation, for the full justification of the mixed economy model it is necessary to establish the percentage value of an optimum threshold for intervention or participation of the state and the public sector in the national economy. This can be done using the econophysics model for the mixed economy grounded and extensively analyzed in this work.

The econophysics model of mixed economy is based on the model of the economic amplifier that works by analogy with an electronic amplifier with electronic amplification devices (transistors or integrated analog circuits) analyzed in our previous papers [4, 9-13]. The econophysics model of the mixed economy proposed in this work is built by analogy with the physical structure of a bipolar junction transistor, widely used in modern microelectronics and digital computing and transmission data etc. The model establishes a methodology whereby the amplification factor of an electronic amplification device $\beta_{el}$ of the junction transistor, in the present case is in direct connection with the $\beta_{ec}$ factor and with the degree of intervention of the State and public sector in the economy, in a wide range of values, represented in a diagram reflecting the correlation between the amplification factor $\beta_{ec}$ and the threshold $\alpha_{ec}$ of State intervention in the economy – i.e. the amount of government expenditure $G$ that intervenes in the economic equilibrium equation for the mixed economy. It is apparent from the analysis of that diagrams that this threshold is lower (approx. 10-25% of GDP) for countries with sustained economic growth, and higher (more than 30%) for countries that through national policies ensure a higher degree of social protection (state pensions, unemployment benefits, free health services and pre-university education etc.).

The decisive success of the mixed economies in the countries that have adopted a low threshold for the degree of interference of the State and its public sector in the economy (the case of the countries of the Neo-American system) constitutes – at the same time – a practical verification of the econophysics model introduced and analyzed in the present paper.

The econophysics model for the mixed economy proposed in this paper is of particular importance for the development of phenomenological econophysics [12] and economic science in general, as well as for its practical application for the development of national economies, taking into account that – as has been shown above – the system of mixed economy is now applied in almost all states, excluding some socialist countries or totalitarian management systems.